\documentclass[useAMS,usenatbib,onecolumn,11pt,a4paper]{article}
\usepackage[figuresright]{rotating}

\usepackage{lscape}
\usepackage{float}
\usepackage[sort&compress]{natbib}
\usepackage{amsmath}
\usepackage{gensymb}
\usepackage{graphicx}
\usepackage{epsfig}
\usepackage{multirow}
\usepackage{setspace}
\usepackage{bigdelim}
\usepackage{bigstrut}
\usepackage{floatflt}	
\usepackage{graphicx}
\usepackage{refstyle}
\usepackage{subcaption}
\usepackage{appendix}
\usepackage[utf8]{inputenc}
\usepackage[export]{adjustbox}  
\usepackage{wrapfig}
\usepackage{pdfpages}

\graphicspath{ {images/} }

\usepackage{caption}

\usepackage[font=Large,labelfont=bf]{caption}
\usepackage{subcaption}
\usepackage[english]{varioref}
\usepackage{lipsum}
\usepackage{hyperref}
\hypersetup{
   colorlinks   = true,
   linkcolor = blue,
   urlcolor= blue,
  citecolor    = blue}

\usepackage{footnote}
\usepackage{multicol}
\usepackage[english]{babel}

\usepackage{blindtext}
 \usepackage{lscape}

\usepackage[compact]{titlesec}
\titlespacing{\section}{0pt}{4ex}{2ex}
\titlespacing{\subsection}{0pt}{1ex}{0ex}
\titlespacing{\subsubsection}{0pt}{0.5ex}{0ex}
\usepackage{indentfirst}

\setcitestyle{square,numbers} 

\labelformat{chapter}{chapter.~#1}
\labelformat{section}{section~#1}
\labelformat{appendix}{app.~#1}
\labelformat{subsection}{section~#1}

\labelformat{figure}{Figure~#1}
\labelformat{subfigure}{figure~\thefigure #1}
\labelformat{table}{Table~#1}
\labelformat{equation}{eq.~(#1)}

\title{{A scalable and reconfigurable industrial-grade Slow Control System for SABRE-South}}

\begin{document}
\maketitle
\flushbottom  
\textit{Authors: Shanti Krishnan$^a$, Scott Collins$^a$, David Smoors$^a$, Craig Webster$^a$, Tiziano Baroncelli$^b$,  Geoffrey Brooks$^a$, Jeremy Mould$^a$, William Joseph Dix$^b$,  Peter Mcnamara$^b$, Federico Scutti$^b$, Greg Lane$^c$,  Phillip Urquijo$^b$, Alan Duffy$^a$.
a: Swinburne University of Technology, John Street, Hawthorn, VIC, Australia
b:  School of Physics, The University of Melbourne, Melbourne, VIC 3010, Australia
c:  Department of Nuclear Physics, Research School of Physical Sciences and Engineering, Australian National University, Canberra ACT2601, Australia}

\pagestyle{plain}


\pagestyle{empty}
\linespread{0.5}

\textit{The Sodium iodide Active Background Rejection Experiment-South (SABRE-South) is a direct dark matter detector soon to be deployed in the Stawell gold mine, in Victoria, Australia.
Monitoring of external environmental and experimental conditions, (temperature, barometric pressure, relative humidity, high voltage, and seismic vibration) is vital to ensure the data quality of the SABRE search for dark matter via direct detection. These parameters have response times ranging from microseconds to seconds and are known as slow control parameters.  
We present the design of a novel compact, industrial-grade, and self-contained slow control system for SABRE-South. This system, featuring innovative hardware and software architecture based on National instruments compact RIO (NI-cRIO) and LabVIEW can be scaled up at low-cost and is capable of implementing the functionalities available in high-end SCADA systems while maintaining the flexibility to integrate custom software code (i.e. C++, PYTHON) for bespoke interfacing needs.}

\section{Introduction}

Sodium iodide Active Background Rejection Experiment-South (SABRE-South), the first ultra-low background direct dark matter experiment in the southern hemisphere will be located in an active gold mine in Stawell, Victoria, Australia \citep{antonello2019sabre}. There is a need to monitor the experimental environment to ensure the data quality of the dark matter particle detection. The environment of the active gold mine is subject to seismic vibrations adding further complexity for maintenance, calibration and health of the detector. 

Large particle physics research centres such as Laboratori Nazionali del Gran Sasso (LNGS) have a well-supported technical infrastructure and use traditional supervisory control and data acquisition (SCADA) system for slow control. An example is the Xenon1T dark matter 
experiment \citep{aprile2017xenon1t} at LNGS. 
We present a self-contained, industrial-grade slow control system using National Instrument's Compact RIO (NI-cRIO) \citep{rioarchitecture}, for SABRE-South, which has a functional performance similar to the traditional SCADA system, can operate reliably in active mine environments and is easily scalable as the experiment evolves. 
The NI-cRIO system is comprised of a real-time controller, reconfigurable input/output (I/O) modules (RIO), field-programmable gate array (FPGA) module and an Ethernet expansion chassis. 
This modular design enables researchers to add a large number of slow control parameters quickly and cost-effectively. The software design allows the storage of a large volume of slow-control data entirely in its local memory for the duration of the life of the experiment, without the need for any additional external servers and ensures the integrity of slow-data in the event of unexpected disruptions. The data is exported periodically to a remote database (i.e. a PostgresSQL server) for monitoring. The system caters to low-speed data acquisition recorded on a scale of seconds (i.e. temperature, barometric pressure, relative humidity) as well as high-speed, resource-intensive data (e.g. vibration) that requires a 5 kHz data rate.

Some of the parameters currently identified for slow control monitoring for SABRE-South include seismic vibrations, high-voltage (HV) supply for the Photomultiplier Tubes (PMTs), temperature, pressure and relative humidity. The design has been set up to monitor local parameters close to the SABRE vessel as well as remote parameters (up to a distance of 100m) through EtherCAT/(TCP/IP) in real-time. The data can be accessed by a remote database. 

In addition, the flexible design allows for the integration of bespoke systems such as muon flux monitors \citep{krishnan2019sipm} for surface condition monitoring of seasonal modulations of cosmic muons, and purpose-built industry-standard power monitoring system to perform reliable long term measurements of the mains power supply.

\begin{figure}
\begin{center}
\includegraphics[width=1.0
   \textwidth]{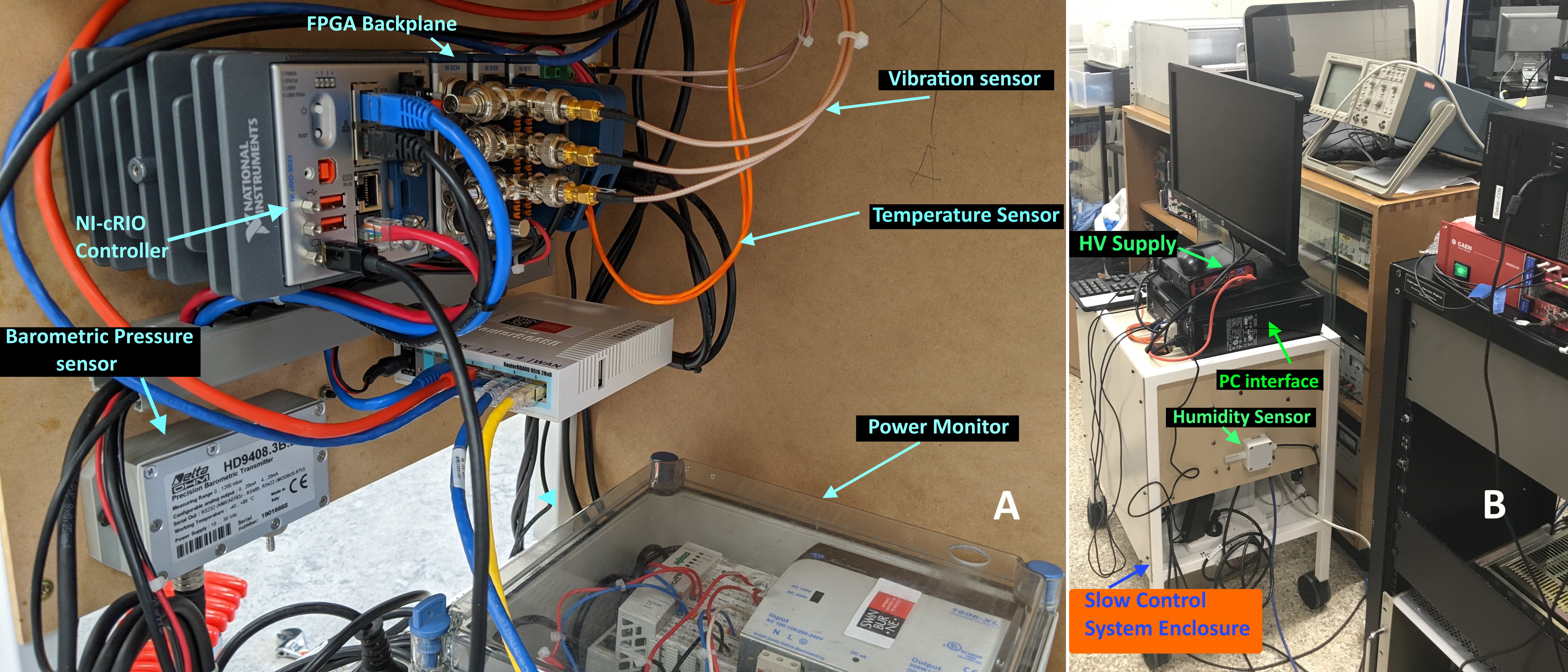}
 \end{center}
\caption{\large{{{ The Slow Control System connected to a three-channel, 0.18m$^2$ muon coincidence system. The system is currently under long-term testing at the University of Melbourne.}}}}
 \label{fig:slow}
\end{figure}

This open-ended and robust system has been operating (since March 2020), in the remote-access mode in the laboratory at the University of Melbourne, monitoring a muon coincidence detector composed of three EJ200 plastic scintillators (each of dimension 60cm x 30cm x 5cm), and read out by Hamamatsu R7724 PMTs, powered by a six-channel VME HV power supply module V6533. This set-up is planned to be deployed to the Stawell gold mine, at depths of 310m and 1025 m below the surface, to measure the muon flux as a function of depth. The slow control set up at the University of Melbourne is depicted in \ref{fig:slow}.

\section{Design Considerations}

One of the primary considerations in this design was to ensure reliable operation in the active mine environment for the lifetime of the experiment, and this mandated industrial-grade performance. For underground experiments such as SABRE that operate without on-site support, the system needs to be able to automatically reboot, in the event of power failure, and handle network disruptions gracefully. The system should be able to handle slow control sensors with different communication interfaces and data collection rates and also have the provision for easy expandability to accommodate additional parameters, as the experiment evolves. The Stawell Underground Physics Laboratory, (SUPL) is newly established, and monitoring requirements may evolve with time. 
The precision and accuracy of the measurement of the environmental and experimental parameters are critical for ensuring the data quality of the dark matter search data and for calibration purposes.
The other key considerations include the ability to perform mixed (analog and digital) measurements, support flexible and effortless adding or replacing of sensors, allow real-time data collection. It should have the ability to adapt to multiple disparate software environments and application interfaces (APIs) and be able to leverage emerging technology trends to meet the current and future needs of the researcher. A mains power supply monitoring system has also been integrated into the slow control design, which could be used to measure the unpredictable mains power in the active mine. 



\section{System Architecture} 
\begin{figure}
\begin{center}
\includegraphics[width=1.0
   \textwidth]{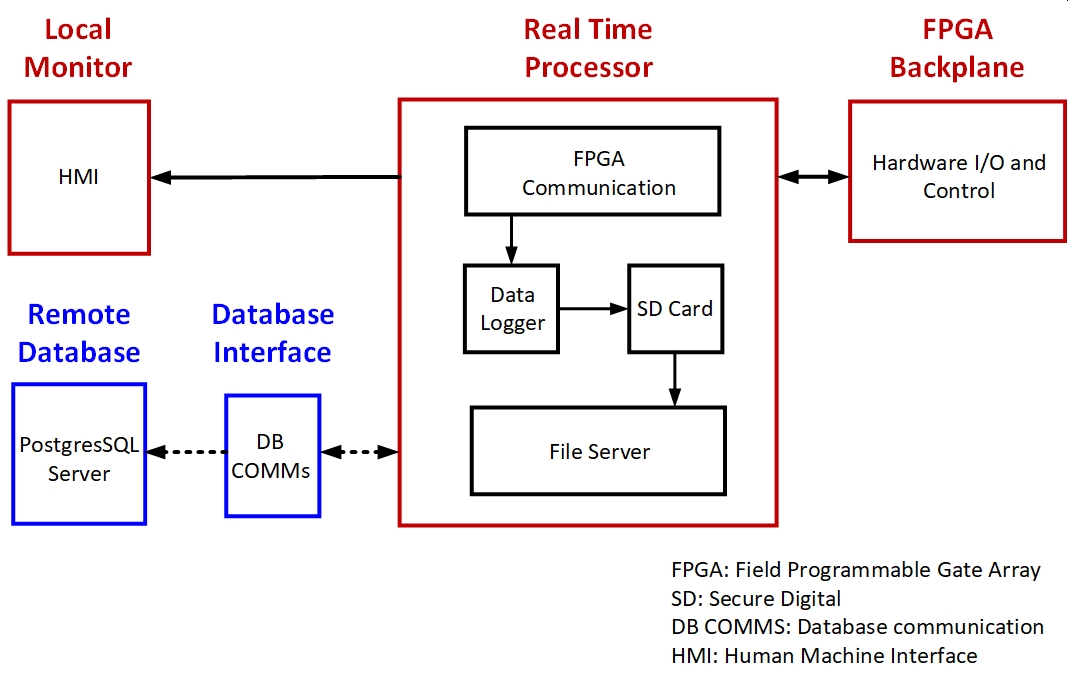}
 \end{center}
\caption{\large{{{Block diagram of the Slow Control System architecture. FPGA is used for data communication between the I/Os and cRIO controller. In some cases, FPGA  performs pre-processing as well.}}}}
 \label{fig:sysarch}
\end{figure}

The central component of the slow control system is a NI-cRIO controller with a processor and a user-programmable FPGA. We chose NI-cRIO, which is an industrial embedded controller system based on the open-source, real-time Linux operating system, as there are proven use-cases of NI-cRIO systems working reliably in
underground environments \citep{rail}.
As shown in \ref{fig:sysarch}, the reconfigurable FPGA backplane of the slow control data logging system collects the data from the I/O modules and communicates it onto the host cRIO real-time processor (RTP). In some cases, such as vibration monitoring, the FPGA pre-processes the data before logging it to the host processor. The real-time processor and FPGA are programmed using LabVIEW (NI LabVIEW) \citep{rioarchitecture} software. The program enables the RTP to store the data on the local onboard memory (SD card) for offline processing and also provide regular, periodic uploads to the remote database. The slow control outputs are displayed on the local human-machine interface (HMI).

\begin{figure}
\begin{center}
\includegraphics[width=1.0
   \textwidth]{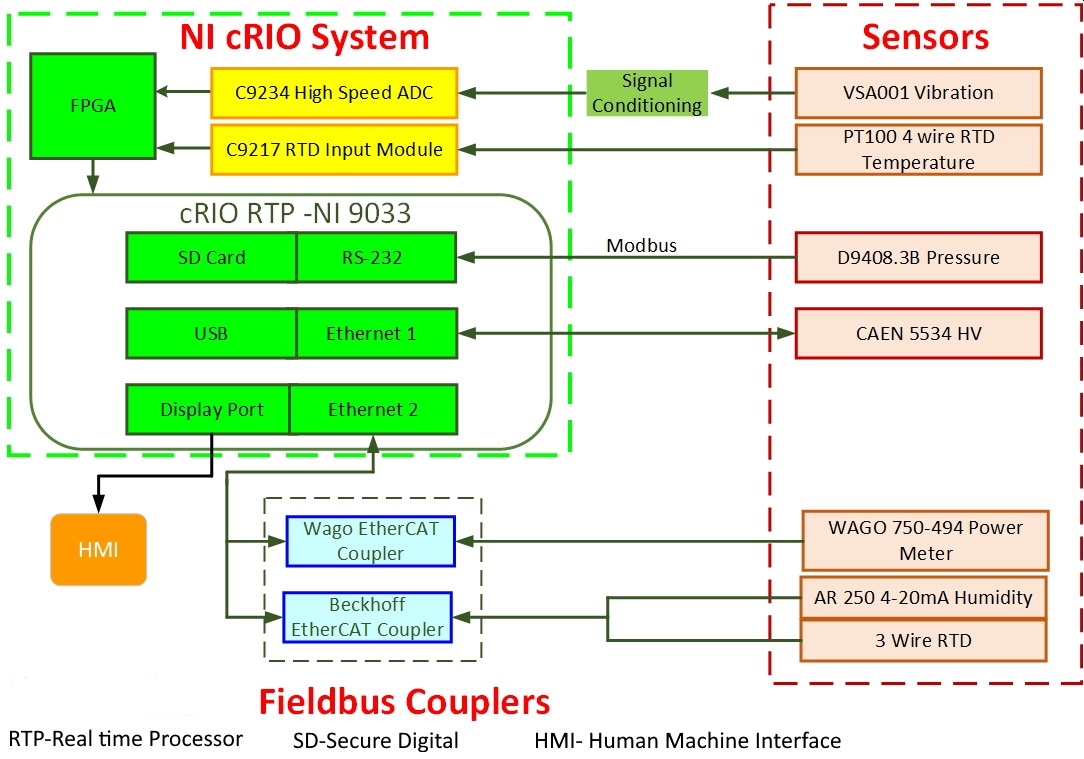}
 \end{center}
\caption{\large{{{
Block diagram displaying a representative integration of the sensors to the cRIO system. It shows the direct integration of the sensors to the I/O modules of the cRIO system as well as remote integration through Fieldbus couplers.}}}}
 \label{fig:sensint}
\end{figure}

As shown in \ref{fig:sensint}, a combination of signal conditioning modules comprising of hot-swappable plug-in I/O modules and hardware controllers are used to interface the sensors to the cRIO controller. The I/O modules are compatible with the cRIO system and feature signal conditioning and conversion circuits. The selection of the industrial sensors has been made to ensure reliable operation in the SUPL environment.
  
\subsection{Integration of Environmental sensors}

Vibration Sensors: Three independent vibration sources in the X, Y and Z directions are measured using three micro-electromechanical systems (MEMS) based accelerometers (IFM-VSA001) \citep{VSA001}, which generate a current output in the range of 0-10 mA. The FPGA sets the vibration event detection condition and performs low-level processing of the high-speed vibration data and streams the data to RTOS for data logging in the onboard memory and local display. 
A terminating resistor converts the current outputs of the vibration sensors to voltage, which is then fed to a 4-Channel, 51.2 kS/s/channel, $+/-$5 V, NI Vibration dynamic signal acquisition input module (NI-9234 \citep{ni9234}), featuring inbuilt antialiasing filters, which allows for automatic adjustment of the sampling rate. The internal delta-sigma ADC of the NI-9234 is configured for a span of 0-5V and a sampling rate of 10.24 KS/s. Application of the Nyquist criterion \citep{ni9234nyq} sets the maximum vibration frequency to $\sim$5 kHz, which meets the performance requirements of SABRE-South experiments and SUPL,  as the maximum expected SUPL vibration frequency is $\sim$1 kHz.  
As shown in \ref{fig:viblavview}, the FPGA pre-processes the input from the sensors, calculates the ${{\text{G}}_{\text{force}}}$ parameters and determines if a valid event has occurred.
${\text{G}}_{\text{force}}$ is calculated by ${{\text{G}}_{\text{force}} = \text{I}/\text{S}}$ where ${{\text{I} = \text{V}_\text{{adc}}/\text{R}_\text{t}}}$, I is the output current of the vibration sensor, S is the sensitivity of the vibration sensor ({${\text{S}}={\text{142}\mu}$A/g) for IFM-VSA001, V$_{\text{adc}}$ is the measured ADC voltage and ${\text{R}}_{\text{t}}$ is the terminating resistor. ${\text{R}}_{\text{t}}$ is chosen to be 500$\ohm$ for an output current range (0 to 10 mA) and positive full span of ADC  (0 to 5V). The resultant vibration amplitude ${\text{G}}_{{\text{sum}}}$ is generated by taking the vector sum of the vibration data in the X,Y and Z direction,  ${\text{G}}_{\text{sum}} = \sqrt({\text{G}}_{\text{x}}^2 +{\text{G}}_{\text{y}}^2+ {\text{G}}_{\text{z}}^2)$. The moving window average ${\text{G}}_{\text{avg}}$ generates a filtered output of the vector sum which is compared to a set threshold and is used to determine valid vibration events. Each sample is bundled up into a data frame of ${\text{G}}_{\text{x}}$, ${\text{G}}_{\text{y}}$, ${\text{G}}_{\text{z}}$, the vector sum amplitude ${\text{G}}_{\text{sum}}$, mean amplitude (moving window average)(${\text{G}}_{\text{avg}}$), and a marker ($'$55$'$) for simple frame validation. 
This data frame is stored in a shared First In First Out (FIFO) buffer in the FPGA, which can be read and processed by the host processor. This parallel and modular approach of using FPGA to pre-process the data reduces the CPU load of the RTOS. LabVIEW has been used to program the FPGA.

\begin{figure}
\begin{center}
\includegraphics[width=1.0
   \textwidth]{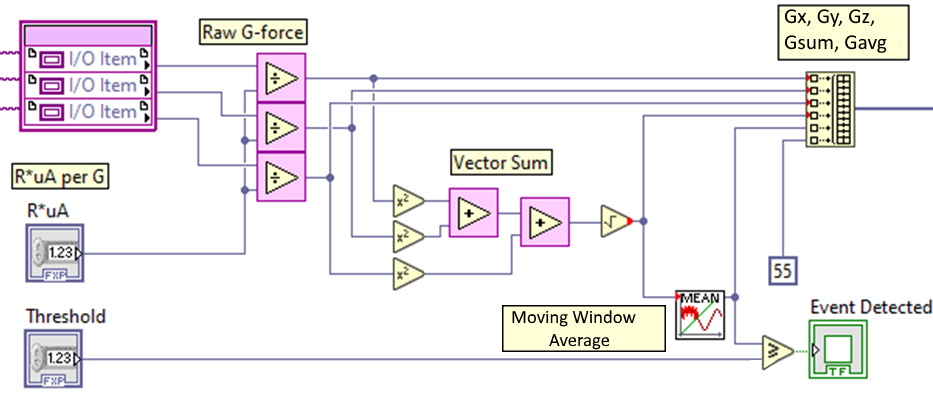}
 \end{center}
\caption{\large{{{ LabVIEW FPGA program segment showing the pre-processing for vibration data before passing to RTOS.}}}}
 \label{fig:viblavview}
\end{figure}

Temperature, barometric pressure and relative humidity sensors: 
Precision RTD temperature sensors have been chosen for the measurement of ambient temperature and the SABRE vessel temperatures. Off the shelf 4-wire PT100 temperature probes (IEC 60751 Class 1/10 : 2008) of accuracy +/- 0.06 deg C  \citep{PT100}, are connected to a NI-9217 \citep{c9217}  RTD analog input module having a temperature accuracy of  +/-0.15$^\circ$C over -40$^\circ$C to +70$^\circ$C.
 
We perform a basic barometric pressure range calculation for SABRE-South laboratory(open air)to specify the upper range of instruments.\\ 
Pressure at average sea-level= 1,013.25 hPa, Stawell Altitude = 230 m, Mine-depth = 1,025 m, SABRE-South is located at a depth of (1025-230 = 795 m) below sea-level. At low altitudes below sea-level, the pressure decreases by $\sim$ 12hPa for every 100 m \citep{miller2009barometric}.
 \begin{align}
 \nonumber\text{Pressure at 795m} &=1013.25+95.4\\ 
  &= 1,108.65 \text{ hPa} \label{eq:pressure}
  \end{align}
  \begin{figure}
\begin{center}
\includegraphics[width=0.6
   \textwidth]{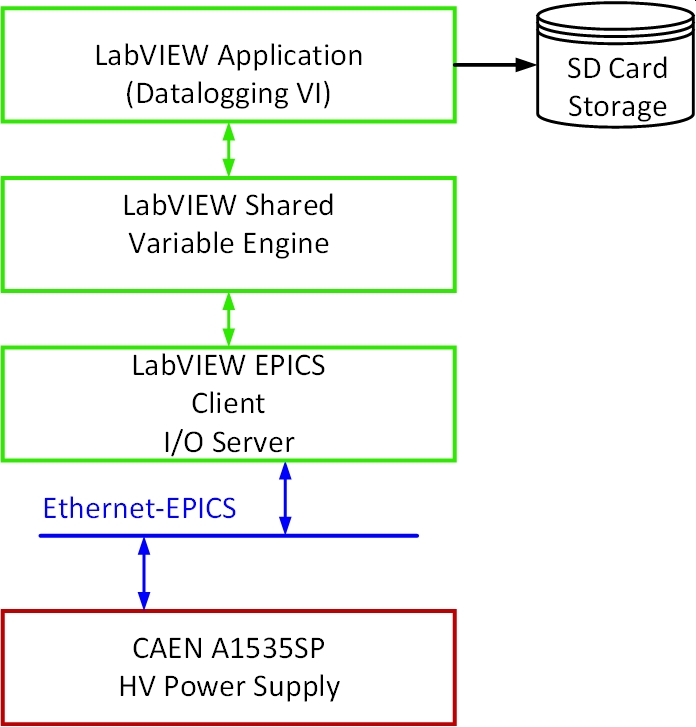}
 \end{center}
\caption{\large{{{Block diagram depicting integration of CAEN A1535SP HV with cRIO using LabVIEW and EPICS.}}}}
 \label{fig:scs}
\end{figure}
For SUPL, as shown in \ref{eq:pressure}, the maximum expected barometric pressure, at a depth of 795m below sea-level, is 1150hPa, with an expected measurement accuracy of  $+/-$ 1hPa stipulated by the SABRE collaboration. To meet these high pressure requirements, high accuracy and reliable piezoresistive sensors Delta Ohm-HD9408.3B \citep{HD9408} capable of measuring pressures from 500 to 1200hPa with operating temperature range -40$^\circ$C  to +85 $^\circ$C range, with a resolution of 0.1hPa has been selected. The barometric transmitter is configured to communicate with the cRIO using standard MODBUS-RTU protocol. 

Relative humidity (RH) transducer APAR-AR250 \citep{AR250} having a measurement accuracy of +\-2$\%$ RH, is integrated to the cRIO system through Beckhoff Ethernet for control automation technology (EtherCAT) couplers \citep{Beckhoff} and uses a 4-20 mA current loop standard.

\subsection{Integration of HV and mains} 
Two types of universal multichannel HV supplies, CAEN A1535SP and CAEN DT 5534 have been integrated into the slow control system.  Each of the HV supplies communicates through different software and connection topologies. CAEN A1535SP HV supplies will be used to power the SABRE vessel PMTs which form a part of the active veto system \citep{bignell2020sabre}.
The HV model CAEN A1535SP supports the open-source software platform, Experimental Physics and Industrial Control Systems (EPICS)\citep{bjorklund2009using}, through its ethernet interface. An EPICS network was used as the interface between CAEN A1535SP HV and the cRIO controller to monitor and control the HV of the PMTs.
As shown in the schematic \ref{fig:scs}, LabVIEW includes an EPICS I/O server-client, which runs directly on the cRIO controller, and functions as the plug-in between the shared LabVIEW process variables and the external network hosting the EPICs devices (EPICS network). The EPICS client I/O server monitors updates to process variables and publishes them to the EPICS network. The system was tested with R5912 PMTs powered by CAEN A1535SP, at the University of Melbourne.
The second HV supply is a CAEN DT 5534 four-channel model, and this unit is planned to be used in the laboratory for R$\&$D work associated with SABRE. As DT 5534 does not support EPICS, it was decided to connect the HV supply via ethernet using TCP communication protocol which allows for remote operation without any additional hardware. LabVIEW application software supported TCP libraries and was used to program the HV supply.
CAEN GECO 2020 software was used to configure the maximum current and voltage settings for CAEN DT 5534 HV and CAEN A1535SP HV, to prevent accidental overvoltage/overcurrent setting by the experimenters, which could result in the damage of the PMTs.

\begin{figure}
\begin{center}
\includegraphics[width=1.0
   \textwidth]{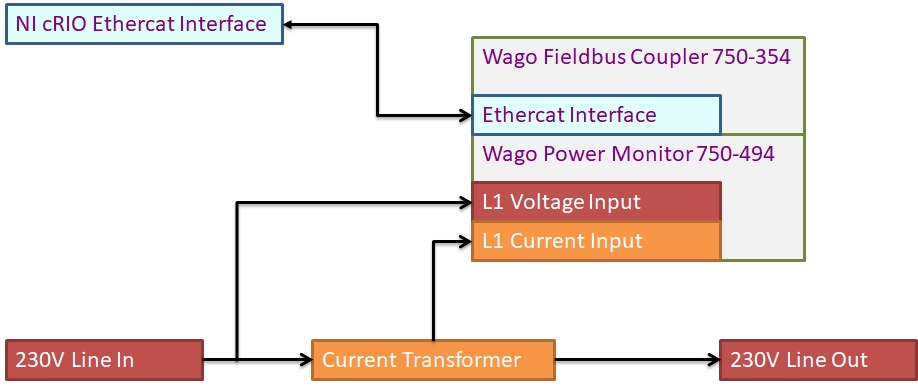}
 \end{center}
\caption{\large{{{Block Diagram of Mains Power Supply monitor (WAGO Power meter) integrated with the EtherCAT fieldbus coupler. }}}}
 \label{fig:wagoethercat}
\end{figure}

The measurement of the mains power supply can be carried out with a three-phase power measurement module, WAGO 750-494 \citep{wago},  assembled in-house using off-the-shelf components. As shown in the block diagram \ref{fig:wagoethercat}, it can measure mains AC voltage, current, frequency, active and reactive power, and is integrated to the slow control system using Wago I/O system EtherCAT Fieldbus coupler 750-354 \citep{wagocoupler}. The module records time-stamped data and rated for 3$\phi$, 480V AC, 1A with a maximum measurement error of $<0.5\%$ for active power.


\section{Implementation of System Software}

LabVIEW is used to program the FPGA and create applications running on cRIO for monitoring and logging slow control data. The FPGA establishes communication between the cRIO and the I/O modules installed in the backplane. 
In most instances, the FPGA transfers the data directly from the I/O modules to the CRIO, with the exception of vibration data, which is pre-processed by the FPGA. LabVIEW has been integrated with EPICS, for monitoring the CAEN high-voltage supplies. 
The communication interface between LabVIEW and EPICS is established, by using an area of shared memory, which allows two relatively independent processes the EPICS server and the LabVIEW runtime to run simultaneously. 

\subsection{ Local and Remote Data Logging}
\begin{figure}
\begin{center}
\includegraphics[width=1.0
   \textwidth]{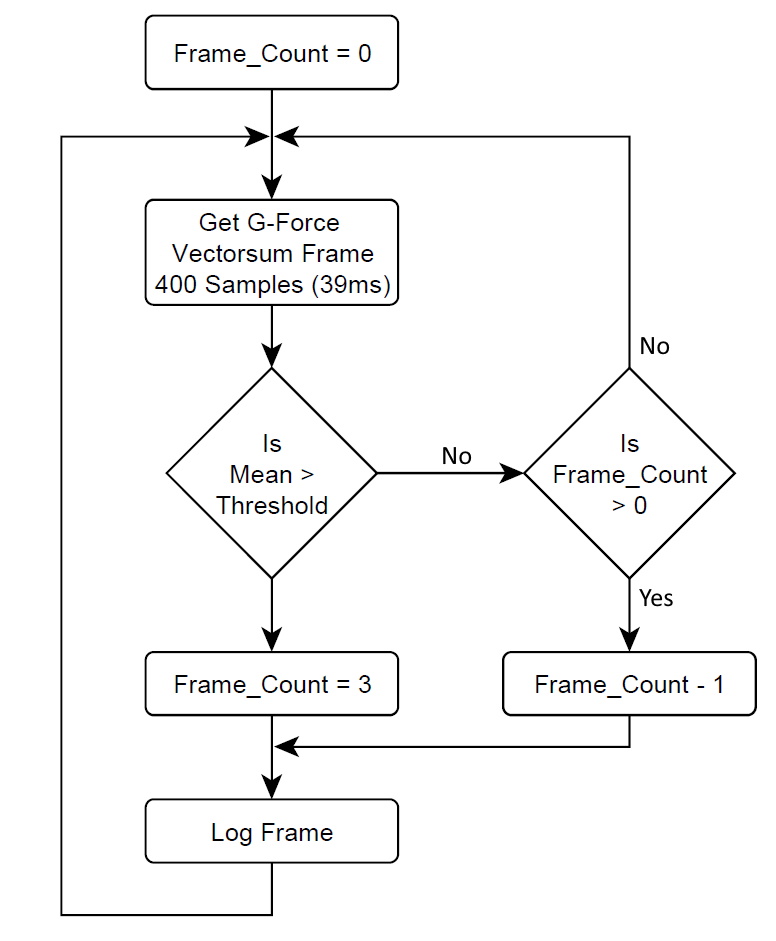}
 \end{center}
\caption{\large{{{Simplified flow-diagram of Vibration Data Logging.}}}}
 \label{fig:vibflow}
\end{figure}

The cRIO host processor (RTP) continuously logs data to three locally stored CSV files. The data include the slow environment data updated every 500 ms (2Hz), periodic logging of valid vibration data, and a log of system events as needed. 

A simplified flow-diagram of vibration data logging is shown in the \ref{fig:vibflow}. 
Vibrations are sampled by the ADC at a sampling rate of 10.24 KS/s and buffered by the FPGA in a 400 sample frame. Every data frame has a duration of 39.2 ms ((1/10.24kS/S)*400 = 39.2 ms). If a valid vibration event is detected by the FPGA, the host processor (cRIO) logs the current data frame and the following three data frames for further processing and storage. If another valid vibration event occurs during the current logging cycle, the logging period is extended to include the following three data frames. The data is stored as comma-separated values (CSV) log file in the cRIO's SD card. By trial and error during the laboratory testing, a data frame of 400 samples (i.e. 39.2 ms) was found to be optimal to minimise CPU load and also have a high probability of recording the events in the presence of noise.  

Environment data is logged in two CSV files, one for vibration and the second for remaining environment data (i.e. temperature, barometric pressure and relative humidity). At the end of four hours of data collection, the two files are compressed into a zip file, and two new CSV log files are created. This periodic zipping operation of four hours results in saving of memory. It also minimises RTP load as the zipping process is time-intensive and the time taken is proportional to the size of data being compressed. In the event of a system reboot/restart, the cRIO controller first checks for incomplete CSV files and zips and logs them before logging new data. This ensures that all the data is captured in the event of an unscheduled shut-down of the system. The size of the vibration CSV file is also continuously monitored, and if it exceeds 500 MB earlier than the periodic time of four hours, the file is zipped, and a new one started. The zipping process compresses the CSV files by around 80-90$\%$.
The third CSV file is used to keep a log of the system status. 

An external interface is required to connect the Linux based cRIO system and the remote PostgresSQL database as currently there is no native NI solution for direct connectivity. Initially, we used a Windows PC, as this was the suggested solution from NI and there was a need to expedite the installation of the slow control system at the University of Melbourne for long-term testing. The Windows operating systems with Web Distributed Authoring and Versioning (WebDAV), an extension of the Hypertext Transfer Protocol (HTTP), was used for interfacing between cRIO and the remote PostgresSQL database, with the bridging program written in LabVIEW. Using a Windows desktop PC / Intel's Next Unit of Computing (Intel-Nuc), introduced additional failure points such as administrative issues on rebooting, power ($\sim$300 Watts) and stability, and this complex solution for a simple interfacing task was ultimately deemed unnecessary. 
\begin{figure}
\begin{center}
\includegraphics[width=1.0
   \textwidth]{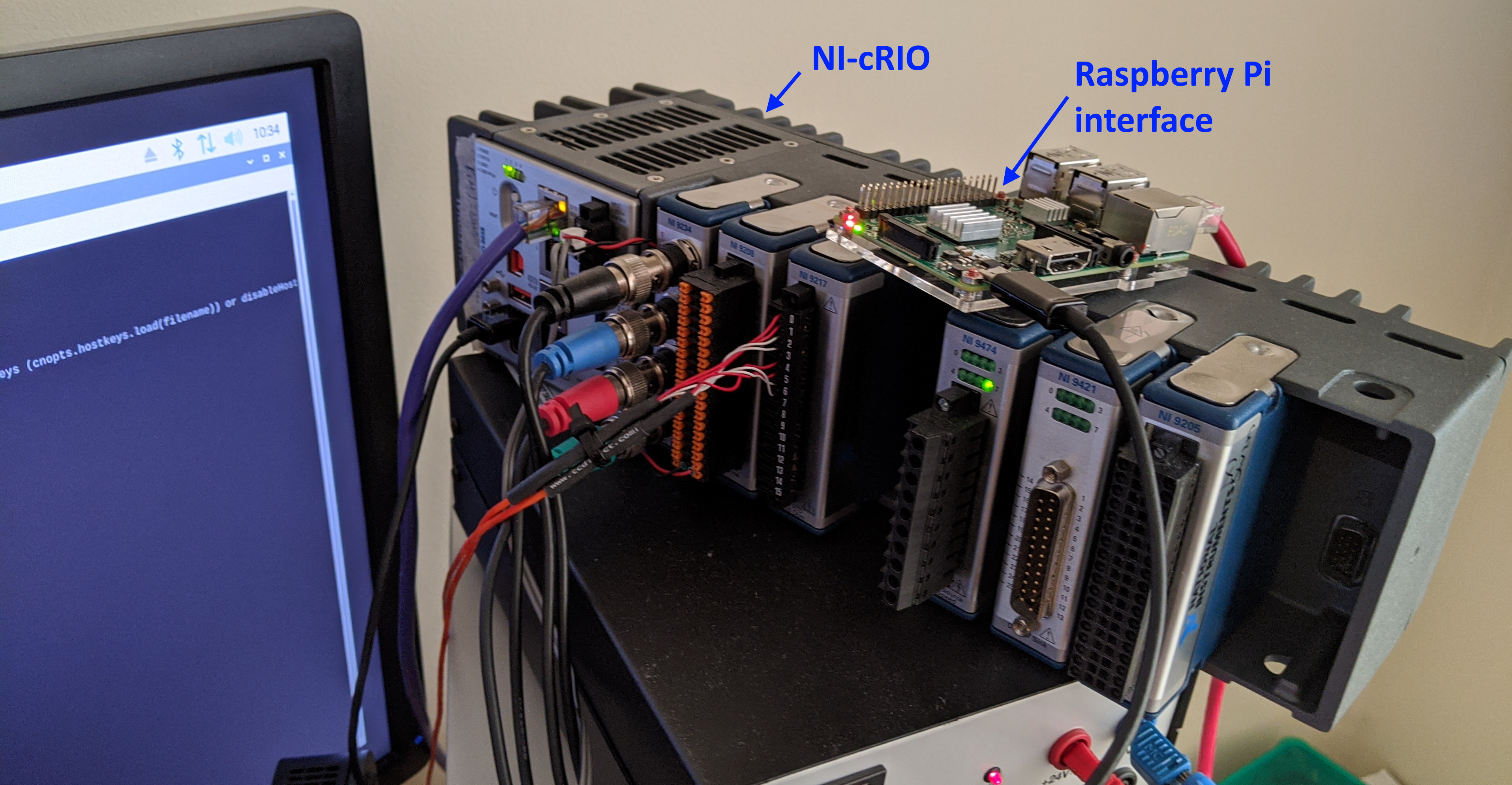}
 \end{center}
\caption{\large{{{ Pre-deployment set-up of the NI-cRIO to Raspberry Pi database interface. }}}}
 \label{fig:riodatabase}
\end{figure}
\begin{figure}
\begin{center}
\includegraphics[width=1.0
   \textwidth]{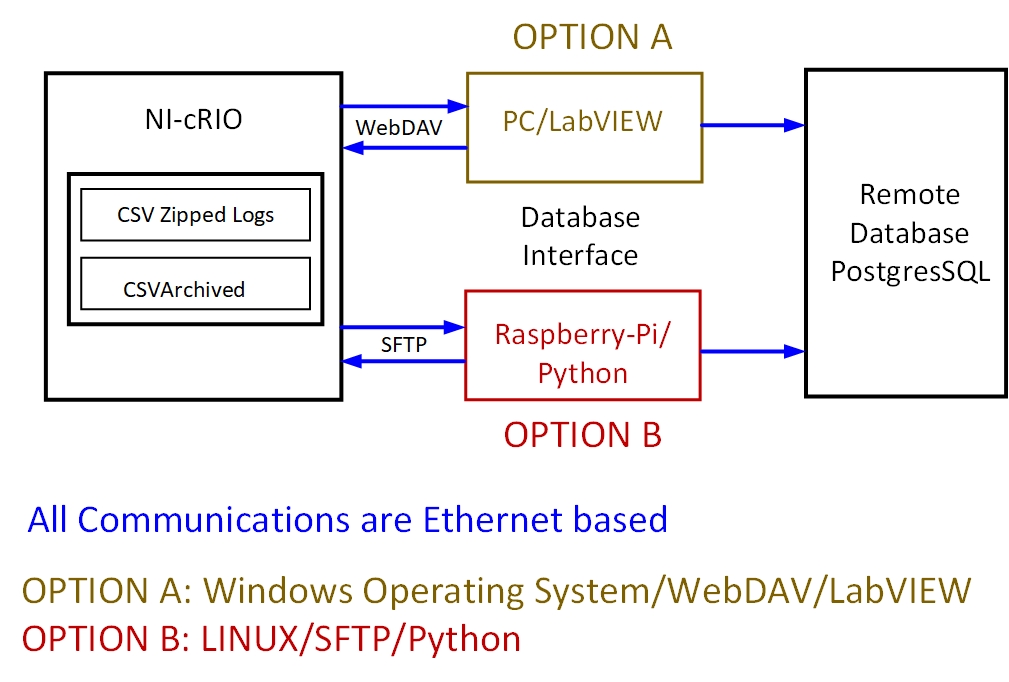}
 \end{center}
\caption{\large{{{Block diagram depicting the bridging interface between Ni-cRIO and remote PostgresSQL database. Option A is with windows PC, WebDAV and Lab-VIEW. Option B is with Raspberry Pi, SFTP protocol and Python. }}}}
 \label{fig:pidatabase}
\end{figure}
We explored options to implement a simple, compact, and energy-efficient solution using a more portable programming language. We have tested a design using Raspberry Pi 3 Model B\citep{rpi}($<$ 10Watts) as shown in \ref{fig:riodatabase}. Secure File Transfer Protocol (SFTP) was used as the interface protocol to access the files (zipped CSV) stored in cRIO. The CSV files were unzipped and the contents transferred to the database. LINUX/ SFTP was used for interfacing between cRIO and the remote PostgresSQL database, with the bridging program written in Python. 
After the completion of CSV data transfer to the database, cRIO receives a command from the database-interface through SFTP and transfers the files to the archive folder of the SD card. The block diagram depicting the two methods of the interface is shown in \ref{fig:pidatabase}. The data transfer rate is slower with the RaspberryPi than with the PC, as shown in \ref{tab:mydb-table}, due to lesser processing power and use of different application programs. However, the data is only transferred to the remote database once every four hours, so this is not a limitation. 
 For use in the mine environment, industrialised Raspberry Pi platform housed in an industrial-grade enclosure and a hardened board design  \citep{industryrpi}, could be used. As Python is easily portable across multiple platforms, Raspberry Pi can be easily switched with other devices such as industrial PC, if necessary.

\begin{table}[]
\centering
\caption{{\small{ Comparison of rate of data-transfer to PostgresSQL database using PC and Raspberry PI. In the table Slow-data refers to Temperature, Barometric pressure and Relative Humidity; PC refers to PC/WebDAV/LabVIEW and Raspberry Pi refers to RaspberryPi/SFTP/Python.}}}
\label{tab:mydb-table}
\begin{tabular}{|l|c|c|c|}
\hline
\multicolumn{1}{|c|}{Log-File} & Sample Data-size      & \multicolumn{2}{c|}{ Data-Transfer Time}                       \\ \hline
                               & \multicolumn{1}{l|}{} & \multicolumn{1}{l|}{PC} & \multicolumn{1}{l|}{Raspberry Pi} \\ \hline
Slow-data & 4.9 MB & 0.311 sec   & 0.556 Sec   \\ \hline
Vibration  & 1.5 MB & 276.1 sec  & 419.41 sec  \\ \hline
\end{tabular}
\end{table}

\section{Results}
Representative results obtained with the slow control system during testing are discussed in this section. As shown in \ref{fig:slvib}, the raw G-force vector sum represents the vibration amplitude and is logged in the cRIO onboard memory, while the filtered moving window average G-force is used as the source to generate an event detection threshold. 
Currently, the maximum vibration data observed over 4 hours of the stress test was 738 kB. At this data rate, a 1GB SD card (currently being used) is capable of storing zipped data up to 255 days. It is possible to store data for the life of the experiment by using a higher capacity SD card. The cRIO is currently configured to delete the oldest zip file if the SD card memory is full. However, it can be configured to perform periodic purging. The system log file tracks the time of deletion of any log files, cRIO restart/reboot time, and any onboard memory FIFO errors.

\begin{figure}
\begin{center}
\includegraphics[width=1.0
   \textwidth]{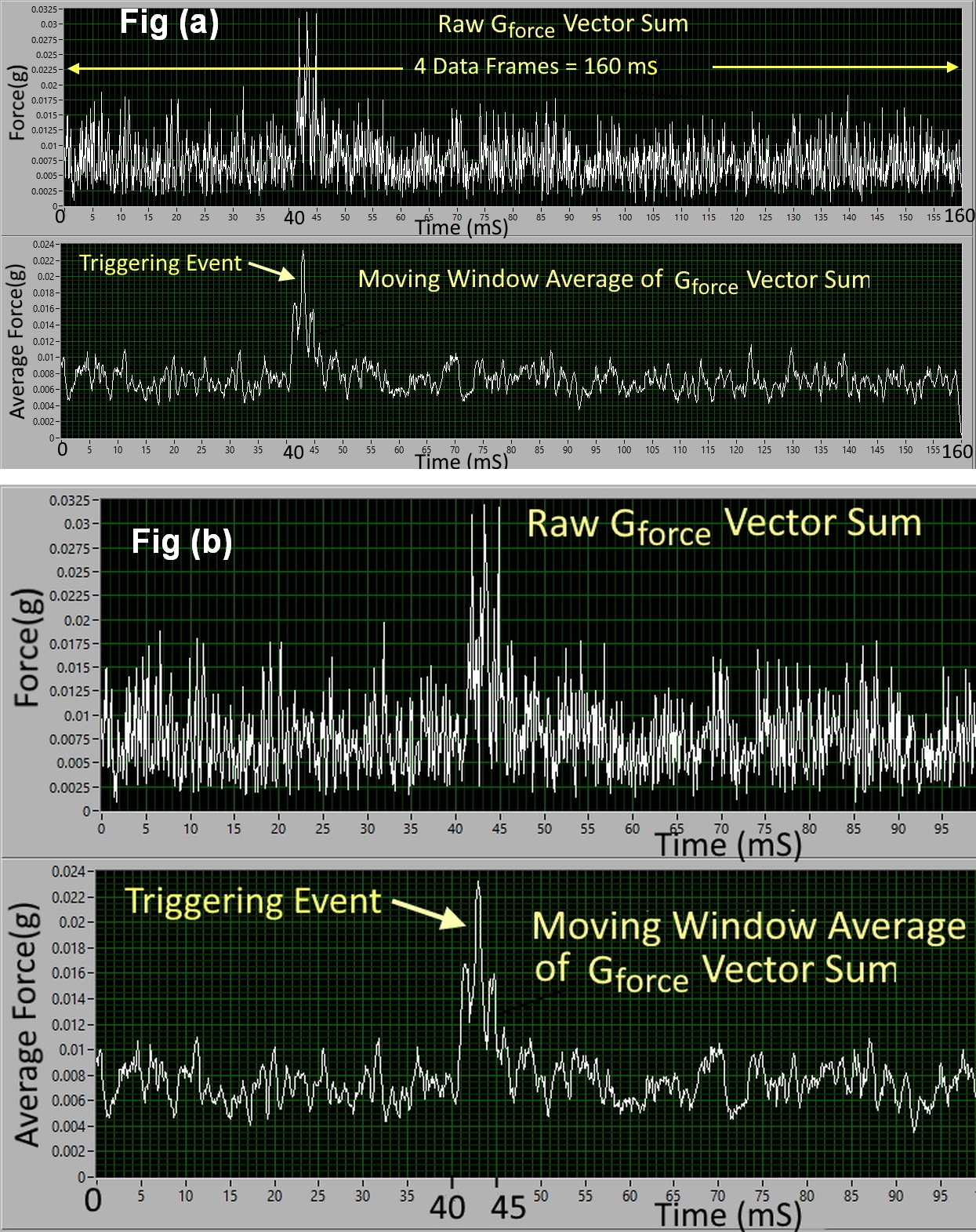}
 \end{center}
\caption{\large{{{Representative graph showing the G$_{\text{force}}$ vector sum and its moving window average for the tri-axial vibration sensor. A tri-axial sensor was mounted on a structural member of the building at Swinburne University and data was generated by light tapping of hammer on the building. Fig (a) shows the data logging of 4 consecutive data frames of the the G$_{\text{force}}$ vector sum and moving window average of the G$_{\text{force}}$ vector sum in case of a single valid vibration event. 
Fig (b) shows the magnitude vs time of occurrence of the G$_{\text{force}}$ vector sum and moving window average of the G$_{\text{force}}$ vector sum.}}}}
 \label{fig:slvib}
\end{figure}
As shown in \ref{fig:HV}, the slow control system displays the independently controlled HV of the four-channel DT 5534 HV module. In the \ref{fig:HV}, the pre-configured settings in channel 3 override user settings and protect the PMTs from accidental over-voltage.  A representative time-stamped data logging CSV output is shown in \ref{tab:my-table}.

\begin{figure}
\begin{center}
\includegraphics[width=1.0
   \textwidth]{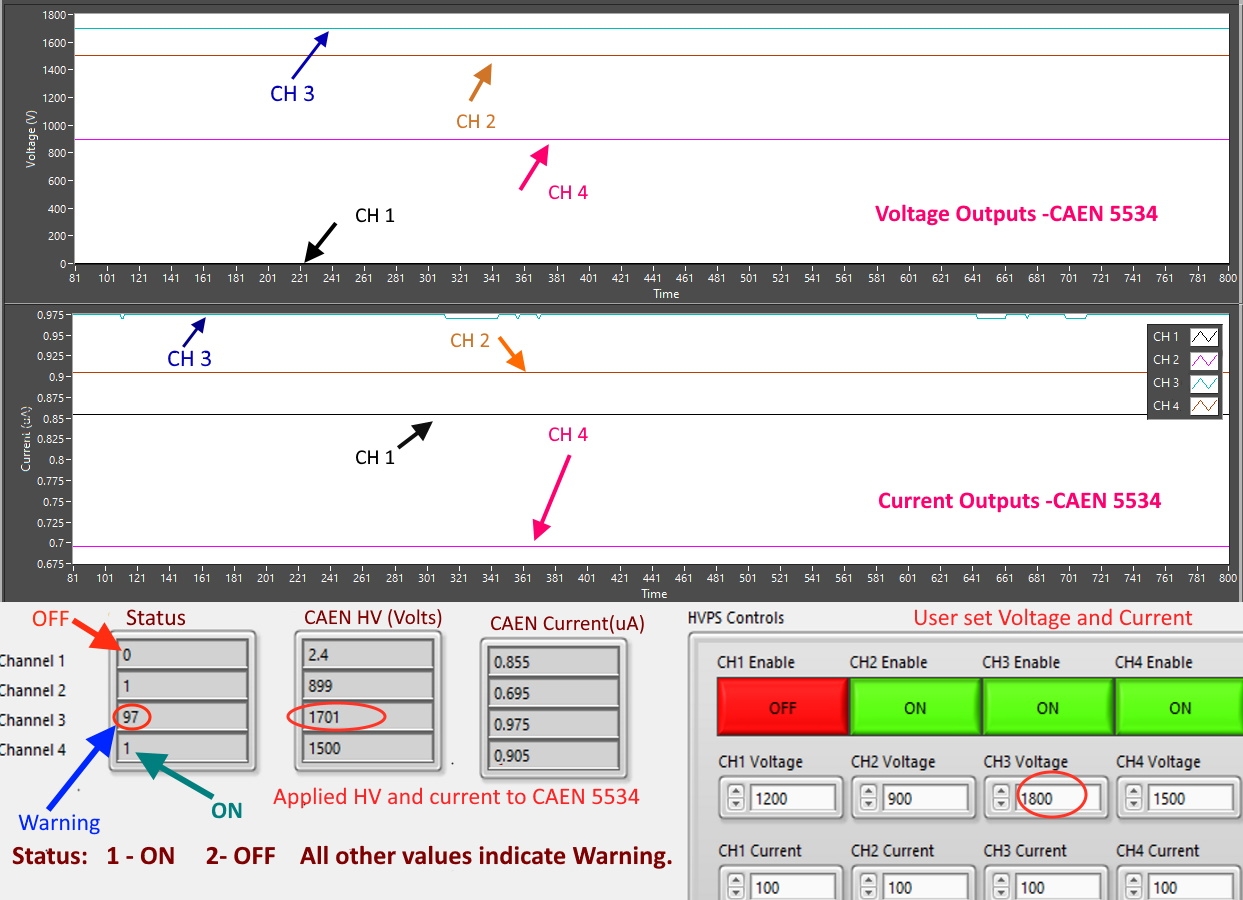}
 \end{center}
\caption{\large{{{Screen shot of the HV control panel showing the historic and current values  for CAEN DT 5534 four channel HV. An example warning status is shown in CH3. In this instance, a warning is issued when the user-set value (1800V) exceeds the preset HV values (1700V) which is set during pre-commissioning using GECO2020. }}}}
 \label{fig:HV}
\end{figure}
\begin{table}[]
\caption{{{ Representative time-stamped data stored as CSV log files in local SD card of cRIO.}}}
\label{tab:my-table}
{\textbf\large{Sample Environmental Data}}\\
\begin{tabular}{|l|l|l|l|l|}
\hline
\begin{tabular}[c]{@{}l@{}}Time\\ Date  \\ 2020-02-24\end{tabular} &
  \begin{tabular}[c]{@{}l@{}}Pressure \\ mbar\end{tabular} &
  \begin{tabular}[c]{@{}l@{}}Humidity\\ \%\end{tabular} &
  RTD1 &
  RTD2 \\ \hline
06\_36\_20.222 & 1014.1 & 59.8  & 25.4 & 24.8  \\ \hline
06\_36\_20.721 & 1014.1 & 59.05 & 25.4 & 24.8  \\ \hline
6\_36\_21.421  & 1014.1 & 58.86 & 25.4 & 24.8  \\ \hline
06\_36\_23.421 & 1014.1 & 58.87 & 25.4 & 24.8  \\ \hline
\end{tabular} \\ \\ \\
{\textbf\large{Sample Vibration Data}}\\
\begin{tabular}{|l|l|l|l|l|}
\hline
\begin{tabular}[c]{@{}l@{}}Time\\ Date  \\ 2020-02-24\end{tabular} & Event & Threshold & Peak & Vsum \\ \hline
58:53.6 & 1 & 0.0135 & 0 & 0.007442 \\ \hline
        &   &        &   & 0.007139 \\ \hline
        &   &        &   & 0.003263 \\ \hline
        &   &        &   & 0.008699 \\ \hline
\end{tabular}\\ \\ \\
{\textbf\large{Sample High Voltage (HVPS) and Mains power (L1) Data}}\\
\begin{tabular}{|l|l|l|l|l|l|l|l|}
\hline
\begin{tabular}[c]{@{}l@{}}Time\\ Date  \\ 2020-02-24\end{tabular} &
  L1 Hz &
  \begin{tabular}[c]{@{}l@{}}HVPS\\ CH 3\end{tabular} &
  \begin{tabular}[c]{@{}l@{}}HVPS\\ CH1\end{tabular} &
  \begin{tabular}[c]{@{}l@{}}HVPS\\ CH4\end{tabular} &
  L1  V &
  L1 W &
  L1 VA \\ \hline
06\_36\_20.222 & 49.98 & 1701.15 & 899.99 & 1500.55 & 248.15 & 50.019 & 48.88 \\ \hline
06\_36\_20.721 & 49.11 & 1701.15 & 899.99 & 1500.55 & 247.97 & 50.019 & 49.11 \\ \hline
6\_36\_21.421  & 49.02 & 1701.15 & 899.99 & 1500.55 & 247.93 & 50.019 & 49.02 \\ \hline
06\_36\_23.421 & 49.51 & 1701.15 & 899.99 & 1500.55 & 247.93 & 50.019 & 49.02 \\ \hline
\end{tabular}\\ 
\end{table}

\begin{figure}
\begin{center}
\includegraphics[width=1.0
   \textwidth]{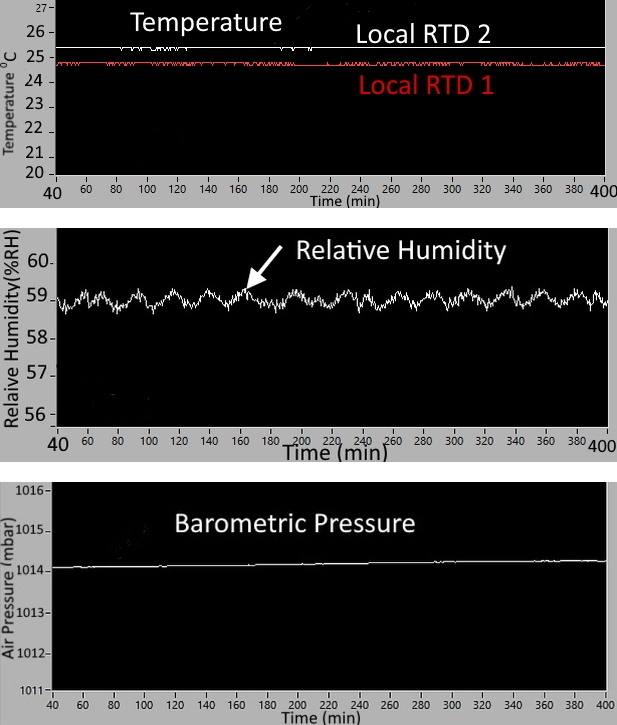}
 \end{center}
\caption{\large{{{Representation of the historic values of temperature, relative humidity and barometric pressure recorded over a period of 6 hours.}}}}
 \label{fig:environment1}
\end{figure}
\begin{figure}
\begin{center}
\includegraphics[width=1.0
   \textwidth]{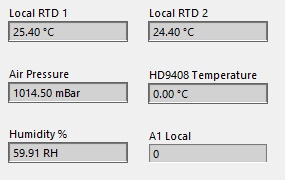}
 \end{center}
\caption{\large{{{Representation of the current values of temperature, barometric pressure and relative humidity.
}}}}
 \label{fig:environment2}
\end{figure}
\begin{figure}
\begin{center}
\includegraphics[width=1.0
   \textwidth]{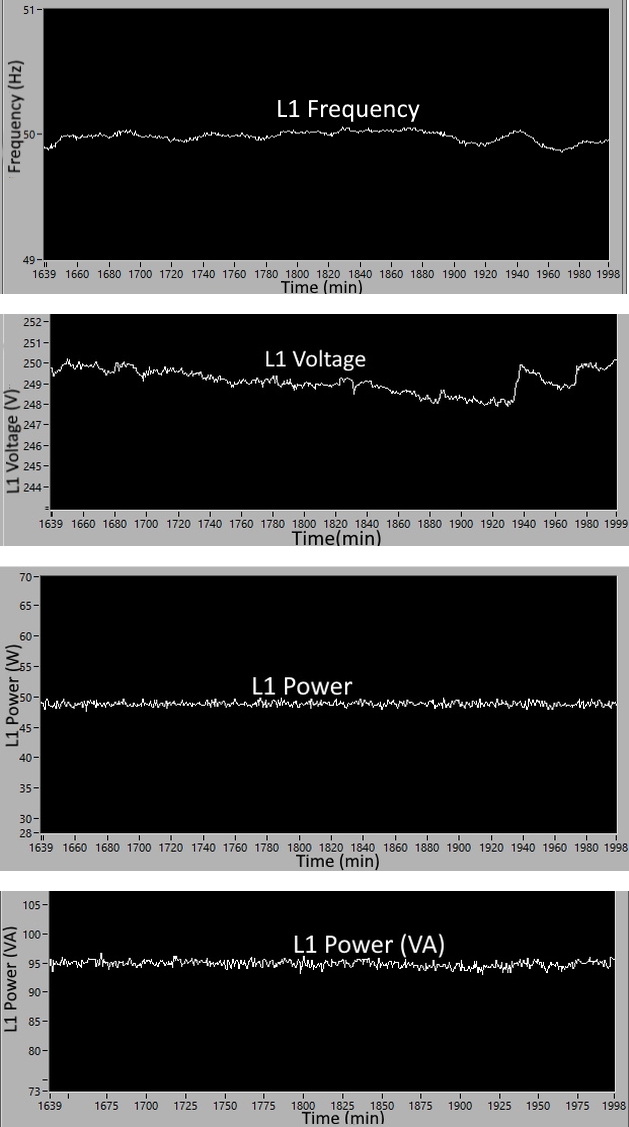}
 \end{center}
\caption{\large{{{Representative Mains power supply monitoring screen showing historic values of mains voltage, frequency and power. }}}}
 \label{fig:power1}
\end{figure}
 
\begin{figure}
\begin{center}
\includegraphics[width=1.0
   \textwidth]{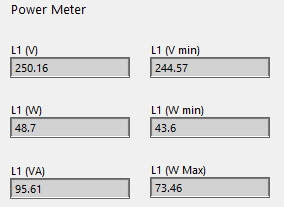}
 \end{center}
\caption{\large{{{Representative Mains power supply monitoring screen showing current values of mains voltage, frequency and power. }}}}
 \label{fig:power2}
\end{figure}

The monitored value of temperature, barometric pressure and relative humidity is shown in \ref{fig:environment1} and \ref{fig:environment2}. The slow control system handles local (i.e.directly connected) and remote (i.e.connected through EtherCAT) sensors interchangeably, without additional configuration.
A representation of the power meter display is shown in \ref{fig:power1} and \ref{fig:power2}.

\section{ Initial Validation of the Design}

We are currently testing the slow control system with a three-channel muon coincidence system at the University of Melbourne.  
The set up has been functioning reliably in an unstaffed environment (due to COVID-19 conditions) since 23$^{{\text{rd}}}$ March 2020. We are monitoring the temperature, barometric pressure, relative humidity and the environmental vibrations in the laboratory through a remote database.
There were network connectivity issues which caused a disruption in data transmission to the remote database from 26$^{{\text{th}}}$ March to 29$^{{\text{th}}}$ April 2020. However, the onboard memory of the cRIO system is configured to keep a backup of all the slow control data. As a result, when network communication was restored,  the system was able to automatically re-synchronise the data collected over this period with the external database. A representative sample of the acquired slow control data is shown in \ref{fig:meldata}. The planned deployment of the slow control system to the Stawell mine, and long-term testing to cross-validate the slow control system data against the standard real-time mine environmental data, will be undertaken when lockdown restrictions are relaxed.

\begin{figure}
\begin{center}
\includegraphics[width=1.0
   \textwidth]{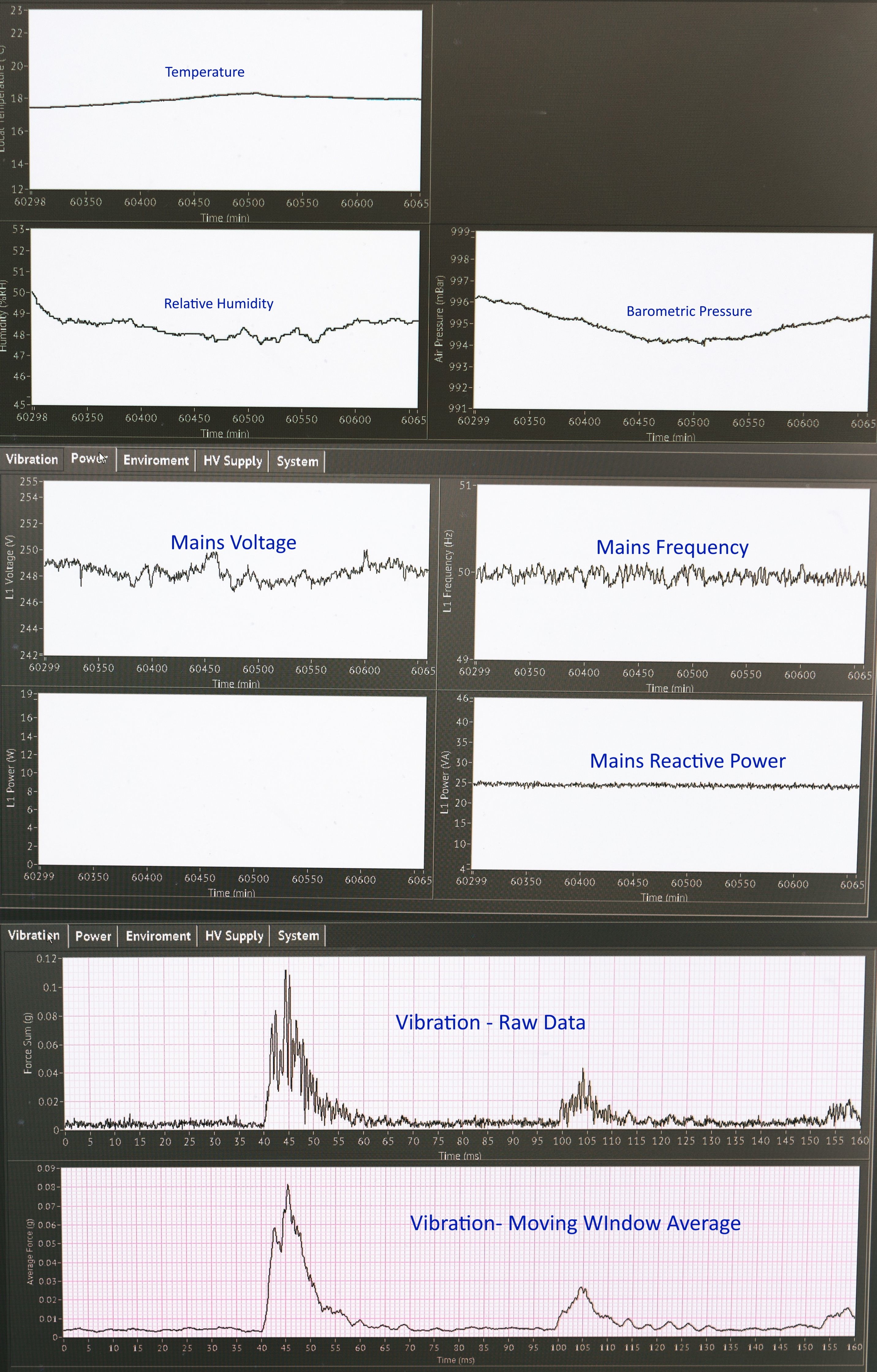}
 \end{center}
\caption{\large{{{Representative Screen shot of Slow Control Data logged at the University of Melbourne (
26$^{th}$ March to 2$^{nd}$ May, 2020. )}}}}
 \label{fig:meldata}
\end{figure}

\section{Proposed SABRE-South Slow Control Considerations}

The SABRE experiment needs to be sensitive enough to reliably measure dark matter search data in the presence of background.

As shown in \ref{fig:sabvess}, the SABRE-South vessel will house a NaI(Tl) detector array having seven crystals, and fourteen PMTs, one on each end of the crystal, and a further 18 PMTs mounted in the active liquid scintillator veto system, 
Linear Alkyl Benzene (LAB). 
Since the PMT dark rates are highly susceptible to temperature variations \citep{Photomultiplier}, this may affect the detection thresholds and hence result in more detection of background. It is therefore critical to monitor and record the short term (hours) and long term (days) averages of the environmental temperature in SABRE which is expected to be in the temperature range of  19$^\circ$C +/- 2$^\circ$C. 
To meet the above requirements, the environmental temperature monit\begin{figure}
\begin{center}
\includegraphics[width= 1.0\textwidth]{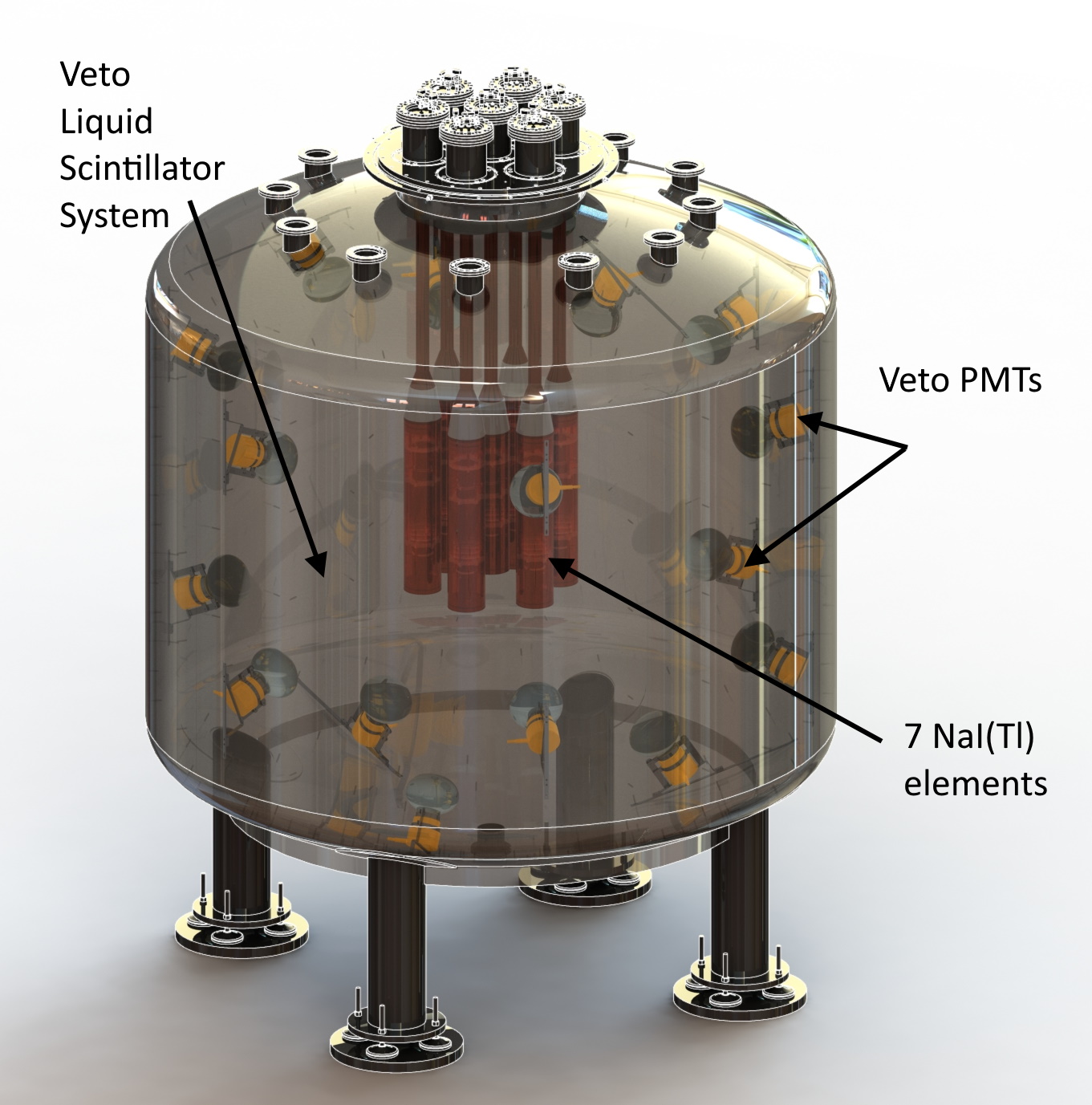}
\caption{{\large Planned SABRE-South Direct Dark Matter Detector showing the detector modules comprising of 7 NaI(Tl)crystals and associated PMTs in Copper tubes, and 18 veto Photomultiplier Tubes (PMTs)immersed in  $\sim$11.6 m$^3$ of Linear Alkyl Benzene(LAB) based veto liquid scintillator.  \textit{{Courtsey: Tiziano Baranocelli, The University of Melbourne}}}}
\label{fig:sabvess}
\end{center}
\end{figure}oring system should have a resolution of the order of 0.2$^\circ$C \citep{calibration}. 

The slow control system is currently able to record the environmental temperature with an accuracy of +/- 0.15$^\circ$C through a combination of precision resistance temperature detector (RTD) temperature probes (PT100), cRIO system compatible RTD analog input module with 24-bit resolution (NI 9217), and the implementation of precision fixed-point arithmetic in the high-speed FPGA \citep{xu2009fpga}.
 A representation of the proposed slow control instrumentation for SABRE-South experiment is depicted in \ref{fig:sabrslo}.
\begin{figure}
\begin{center}
\includegraphics[width=1.0
   \textwidth]{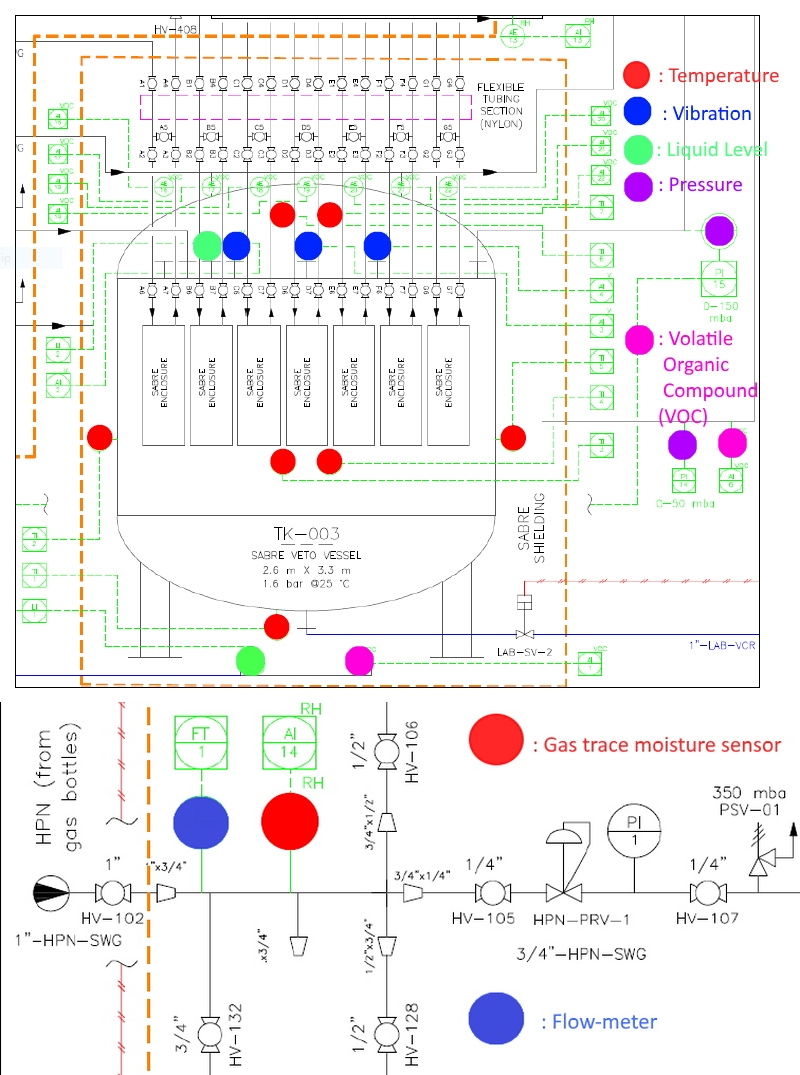}
 \end{center}
\caption{\large{{{Representation of the proposed slow control instrumentation for SABRE-South. \textit{{Courtsey: Tiziano Baranocelli, The University of Melbourne}}}}}}
 \label{fig:sabrslo}
\end{figure} 
The SABRE-South vessel filled with $\sim$11.6 m$^3$ of LAB would be expected to reach thermal equilibrium with the environment. 
The inclusion of PMTs and other instrumentation will result in additional heat flow into the SABRE vessel, causing the vessel to be at a higher temperature than the SABRE laboratory as has been observed in other experiments such as COSINE-100 \citep{adhikari2018initial}. 
At this stage, the magnitude of the thermal difference is not known. The SABRE vessel is not provided with any substantive active cooling mechanism and relies upon the stability of the air-conditioning of the surrounding environment, i.e. the laboratory, as the major driver for the thermal equilibrium temperature. 
The temperature monitoring system of the SABRE vessel is expected to have an accuracy of $<<$ 0.2$^\circ$C. It is possible to achieve this accuracy by designing a bespoke system spanning a much narrower range of temperatures (i.e. 20$^\circ$C) and calibrating it.

High relative humidity could negatively impact the function, stability, and the life of the electronics equipment and must be monitored for safe operation \citep{khanna2017extreme}. For testing purposes, an industrial-grade relative humidity sensor with a tolerance of $\pm 2\%$ tolerance has been used and can be easily replaced with meteorological grade RH sensor such as Vaisala-Hmt310 \citep{HMT310}, having $\pm 1 \%$ tolerance if required, as both sensors have standard 4-20mA current outputs. The air-conditioning system will maintain the relative humidity of the SABRE laboratory in the 30$\%$-50$\%$ range.

When subjected to external vibrations and shocks, PMTs are known to exhibit changes in signal levels and microphonic noise, and excessive vibrations could result in their breakage \citep{Photomultiplier}. As the active mine environment in Stawell will experience vibrations due to heavy machinery operation and explosions for mining excavations, the environmental vibrations must be closely monitored, to minimise the damage to sensitive equipment such as PMTs. As the veto PMTs will be submerged in the LAB, the corresponding response of the PMTs to the changed environmental conditions need to be considered.

One challenge that the Thallium doped Sodium Iodide crystal (NaI(Tl)) presents as a dark-matter target is its highly hygroscopic nature. A crystal must be protected from moisture at all times to prevent degradation of its optical properties. 
The modular detector array has crystals sealed in enclosures, that can be thoroughly outgassed using a continuous flow of high purity dry nitrogen (DN$_2$) to remove trace amounts of water clinging to the surfaces \citep{shields2015sabre}. 
Monitoring the moisture content of (DN$_2$) at the inlet and outlet of the detector array is critical. The slow control system allows for this operation through integration with industry grade sensors such as Pura Pure Gas Trace Moisture Transmitter \citep{moisture} or bespoke design built using I$^2$C Sensirion SHT21 \citep{sensirion}. The  DN$_2$ gas flow-rate,  volatile organic compound (VOC) vapours on N$_2$ exhaust and in the catchment at the base of SABRE vessel to detect LAB leaks, DN$_2$ gas-pressure in the LAB and SABRE vessel shielding, level of LAB fluid, are just some of the additional identified parameters which the slow control system could monitor.

\section{Conclusion}

The prototype slow control system has been designed with a modular architecture to enable scaling by simply daisy-chaining it to other Fieldbus controllers. The system architecture supports different communication protocols (i.e. Ethernet, MODBUS, RS 232, USB, EtherCAT) enabling seamless integration of local and remote sensors (up to a distance of up to 100m) from the experimental set-up and a substantial reduction in space requirements and wiring overheads.

The flexible system design allows the integration of heterogeneous sensors and systems enabling monitoring of parameters such as vibration, cosmic muon flux and mains power.
 The parallel approach of using FPGA for data processing reduces the real-time processor (RTP) load. The direct connectivity of the FPGA with the I/O modules allows for deterministic interface timing which may be required for process-intensive external sensors.

Logging time-stamped data in the onboard memory for redundancy and data back up of the last operation after a power shut-down, ensures the integrity of the system in the event of network communication errors or power outage, which is to be expected in the unpredictable, 
active mine environment. Thus, the system is able to provide continuous and reliable background monitoring despite interruptions in power supply and changes in the environment. 
This versatile system is flexible enough to support both critical industrial sensors and research-based bespoke systems on the other. It enables the consolidation of different types of background data in one platform, enabling ease of analysis, monitoring and control.

This open-ended real-time slow control system design is capable of industrial-grade performance/reliability and is expected to meet the key current and future emerging needs of the experiments in a subterranean environment. 
\section{Acknowledgement}
The authors acknowledge the support received from Swinburne University of technology and the University of Melbourne for making available the equipment and resources to carry out the development and testing.
    \setcitestyle{numbers}
\bibliographystyle{mnras}


{\footnotesize
\setlength{\itemsep}{1pt}
\bibliography{bibliography}{}

\begin{thebibliography}{}
\makeatletter
\relax
\def\mn@urlcharsother{\let\do\@makeother \do\$\do\&\do\#\do\^\do\_\do\%\do\~}
\def\mn@doi{\begingroup\mn@urlcharsother \@ifnextchar [ {\mn@doi@}
  {\mn@doi@[]}}
\def\mn@doi@[#1]#2{\def\@tempa{#1}\ifx\@tempa\@empty \href
  {http://dx.doi.org/#2} {doi:#2}\else \href {http://dx.doi.org/#2} {#1}\fi
  \endgroup}
\def\mn@eprint#1#2{\mn@eprint@#1:#2::\@nil}
\def\mn@eprint@arXiv#1{\href {http://arxiv.org/abs/#1} {{\tt arXiv:#1}}}
\def\mn@eprint@dblp#1{\href {http://dblp.uni-trier.de/rec/bibtex/#1.xml}
  {dblp:#1}}
\def\mn@eprint@#1:#2:#3:#4\@nil{\def\@tempa {#1}\def\@tempb {#2}\def\@tempc
  {#3}\ifx \@tempc \@empty \let \@tempc \@tempb \let \@tempb \@tempa \fi \ifx
  \@tempb \@empty \def\@tempb {arXiv}\fi \@ifundefined
  {mn@eprint@\@tempb}{\@tempb:\@tempc}{\expandafter \expandafter \csname
  mn@eprint@\@tempb\endcsname \expandafter{\@tempc}}}

\bibitem[\protect\citeauthoryear{APAR}{APAR}{2018}]{AR250}
APAR 2018, {Humidity transducer},
  \url{https://www.apar.pl/en/humidity-transducer-ar250.html}

\bibitem[\protect\citeauthoryear{Adhikari et~al.,}{Adhikari
  et~al.}{2018}]{adhikari2018initial}
Adhikari G.,  et~al., 2018, The European Physical Journal C, 78, 1

\bibitem[\protect\citeauthoryear{Antonello et~al.,}{Antonello
  et~al.}{2019}]{antonello2019sabre}
Antonello M.,  et~al., 2019, The European Physical Journal C, 79, 1

\bibitem[\protect\citeauthoryear{Aprile et~al.,}{Aprile
  et~al.}{2017}]{aprile2017xenon1t}
Aprile E.,  et~al., 2017, The European Physical Journal C, 77, 1

\bibitem[\protect\citeauthoryear{Bignell et~al.,}{Bignell
  et~al.}{2020}]{bignell2020sabre}
Bignell L.~J.,  et~al., 2020, in EPJ Web of Conferences. p. 01002

\bibitem[\protect\citeauthoryear{Bj{\"o}rklund, Veeramani  \&
  Debelle}{Bj{\"o}rklund et~al.}{2009}]{bjorklund2009using}
Bj{\"o}rklund E.,  Veeramani A.,   Debelle T.,  2009, in Proceedings of
  ICALEPCS. pp 555--557

\bibitem[\protect\citeauthoryear{Hamamatsu}{Hamamatsu}{2006}]{Photomultiplier}
Hamamatsu 2006, {Photomultiplier tubes-Basics and Applications,3rd Edition},
  \url{https://www.hamamatsu.com/resources/pdf/etd/PMT_handbook_v3aE.pdf//}

\bibitem[\protect\citeauthoryear{IFM}{IFM}{}]{VSA001}
IFM, {VSA001}, \url{https://www.ifm.com/us/en/product/VSA001//}

\bibitem[\protect\citeauthoryear{Instruments}{Instruments}{}]{c9217}
Instruments N., , {Temperature Input Module},
  \url{https://www.ni.com/en-my/support/model.ni-9217.html/}

\bibitem[\protect\citeauthoryear{Khanna}{Khanna}{2017}]{khanna2017extreme}
Khanna V.~K.,  2017, Extreme-temperature and harsh-environment electronics.
IOP Publishing Limited Bristol

\bibitem[\protect\citeauthoryear{Krishnan, Webster, Duffy, Brooks, Clay  \&
  Mould}{Krishnan et~al.}{2019}]{krishnan2019sipm}
Krishnan S.,  Webster C.,  Duffy A.,  Brooks G.,  Clay R.,   Mould J.,  2019,
  Journal of Instrumentation, 14, P09026

\bibitem[\protect\citeauthoryear{MICHELL}{MICHELL}{}]{moisture}
MICHELL, {Pura Pure Gas Trace Moisture Transmitter},
  \url{https://www.michell.com/uk/products/pura.htm//}

\bibitem[\protect\citeauthoryear{Miller, Vandome  \& McBrewster}{Miller
  et~al.}{2009}]{miller2009barometric}
Miller F.,  Vandome A.,   McBrewster J.,  2009, Barometric Formula.
VDM Publishing, \url {https://books.google.com.au/books?id=-p-PQgAACAAJ}

\bibitem[\protect\citeauthoryear{National-Instruments}{National-Instruments}{2018}]{rioarchitecture}
National-Instruments 2018, {The LabVIEW RIO Architecture: A Foundation for
  Innovation}, \url{http://www.ni.com/white-paper/10894/en//}

\bibitem[\protect\citeauthoryear{RS-components}{RS-components}{}]{rpi}
RS-components, {Raspberry Pi 3},
  \url{https://components101.com/sites/default/files/component_datasheet/Raspberry%20Pi%203%20Datasheet.pdf//}

\bibitem[\protect\citeauthoryear{S.Etchell}{S.Etchell}{2012}]{rail}
S.Etchell 2012, {London Underground Improves Reliability for 200 Million Annual
  Passengers with Remote Condition Monitoring},
  \url{http://sine.ni.com/cs/app/doc/p/id/cs-16209#//}

\bibitem[\protect\citeauthoryear{Sensiron}{Sensiron}{}]{sensirion}
Sensiron S., , {Digital Humidity Sensor SHT2x (RH/T)},
  \url{https://www.sensirion.com/en/environmental-sensors/humidity-sensors/humidity-temperature-sensor-sht2x-digital-i2c-accurate///}

\bibitem[\protect\citeauthoryear{Shields}{Shields}{2015}]{shields2015sabre}
Shields E.~K.,  2015, PhD thesis, Princeton University

\bibitem[\protect\citeauthoryear{Vaisala}{Vaisala}{}]{HMT310}
Vaisala, {Humidity transmitter2},
  \url{https://www.vaisala.com/sites/default/files/documents/HMT310-Datasheet-B210769EN-H.pdf}

\bibitem[\protect\citeauthoryear{WAGO}{WAGO}{b}]{wagocoupler}
WAGO, {WAGO EtherCAT coupler},
  \url{https://www.wago.com/au/io-systems/fieldbus-coupler-ethernet/p/750-354}

\bibitem[\protect\citeauthoryear{WAGO}{WAGO}{a}]{wago}
WAGO, {WAGO power measurement},
  \url{https://www.wago.com/au/io-systems/3-phase-power-measurement/p/750-494/}

\bibitem[\protect\citeauthoryear{Xu, Shuang, Jiang  \& Wu}{Xu
  et~al.}{2009}]{xu2009fpga}
Xu Y.,  Shuang K.,  Jiang S.,   Wu X.,  2009, in 2009 International Conference
  on Measuring Technology and Mechatronics Automation. pp 384--387

\bibitem[\protect\citeauthoryear{beckhoff}{beckhoff}{}]{Beckhoff}
beckhoff, {ethercat},
  \url{https://www.ethercat.org/pdf/english/ETG_Brochure_EN.pdf//}

\bibitem[\protect\citeauthoryear{deltaohm}{deltaohm}{}]{HD9408}
deltaohm, {pressure sensor},
  \url{http://www.deltaohm.com/ver2012/download/HD9408_3B_M_uk.pdf}

\bibitem[\protect\citeauthoryear{direct}{direct}{}]{PT100}
direct T., , {Precision RTD},
  \url{https://blog.beamex.com/pt100-temperature-sensor#RTD-sensors}

\bibitem[\protect\citeauthoryear{national instruments}{national
  instruments}{a}]{ni9234}
national instruments, {C Series Sound and Vibration Input Module},
  \url{https://www.ni.com/pdf/manuals/374238a_02.pdf///}

\bibitem[\protect\citeauthoryear{national instruments}{national
  instruments}{b}]{ni9234nyq}
national instruments, {Sound and Vibration data acquisition},
  \url{http://static6.arrow.com/aropdfconversion/8b558ae7357193508d2ec14449ab50300a34a596/cat_wls_923x.pdf/}

\bibitem[\protect\citeauthoryear{netpi}{netpi}{}]{industryrpi}
netpi, {industrial grade raspberry Pi},
  \url{https://www.netiot.com/netpi/industrial-raspberry-pi-3///}

\bibitem[\protect\citeauthoryear{stanford}{stanford}{}]{calibration}
stanford, {Calibration with confidence},
  \url{https://web.stanford.edu/group/csp/cs21/calibration.html//}

\makeatother
\end{thebibliography}

\end{document}